%% file: open_problem.tex
\documentclass[letterpaper]{article}

\usepackage{wafr}
\usepackage{times}
\usepackage{helvet}
\usepackage{courier}
\usepackage{xspace}
\usepackage{color}
\usepackage{hyperref}
\hypersetup{bookmarksopen,bookmarksnumbered,
pdfpagemode=UseOutlines,
colorlinks=true,
linkcolor=blue,
anchorcolor=blue,
citecolor=blue,
filecolor=blue,
menucolor=blue,
urlcolor=blue
}
\usepackage{amsmath,amssymb,amsfonts,amsfonts}
\usepackage{algorithm}
\usepackage[noend]{algorithmic}
\usepackage{url}
\usepackage{graphicx,subfigure}

\usepackage{float}
\usepackage{bm}

\usepackage{multirow}

\title{\LARGE \bf
Open problem on risk-aware planning in the plane
}
\author{
Oren Salzman \and 
Siddhartha Srinivasa \\
The Robotics Institute 
Carnegie Mellon University Pittsburgh, PA
}

\input{macros.tex}

\begin{document}

\maketitle
\thispagestyle{empty}
\pagestyle{empty}

\begin{abstract}
We consider the motion-planning problem of planning a collision-free path of a robot in the presence of risk zones.
The robot is allowed to travel in these zones but is penalized in a super-linear fashion for consecutive accumulative time spent there.
We recently suggested a natural cost function that balances path length and risk-exposure time. 
When no risk zones exists, our problem resorts to computing minimal-length paths which is known to be computationally hard in the number of dimensions.
It is well known that in two-dimensions computing minimal-length paths can be done efficiently.
Thus, a natural question we pose is ``Is our problem computationally hard or not?''
If the problem is hard, we wish to find an approximation algorithm to compute a near-optimal path.
If not, then a polynomial-time algorithm should be found.
\end{abstract}

\section{Introduction}
\label{sec:introduction}

\begin{figure}[tb]
  \centering
  	\includegraphics[height = 4.cm ]{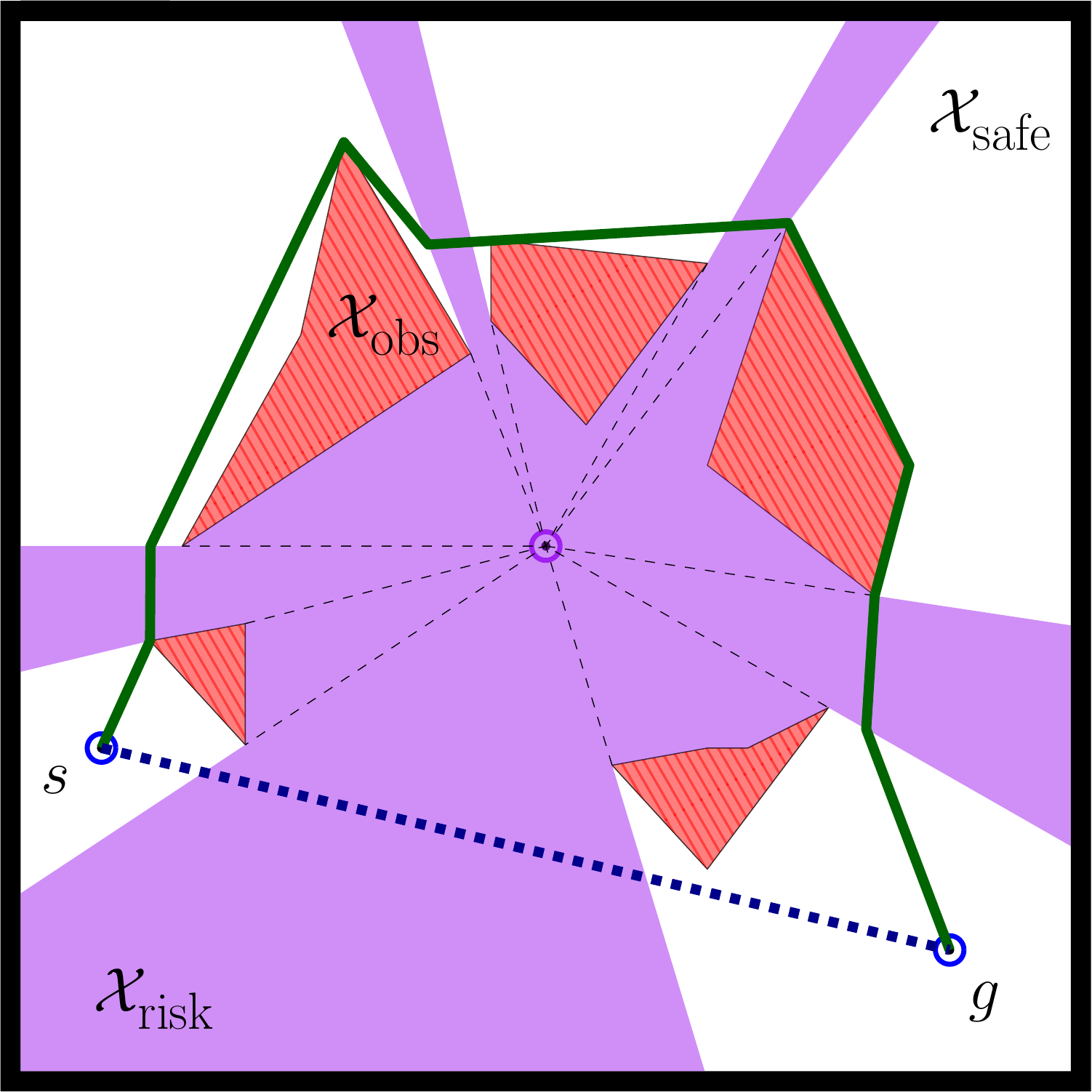}
  \caption{
    \captionstyle
  	Risk aware motion planning.
  	We need to plan a minimal-cost connecting $s$ and $g$ while avoiding obstacles (red). Our cost function penalizes continous exposure to the risk regions (purple), thus the optimal path (solid green) favours intermittent exposure over the long exposure taken by the shortest path (dotted blue). 
  	}
   	\label{fig:example}
	\vspace{-5.5mm}
\end{figure}

We are interested in motion-planning problems where an agent has to compute the least-cost path to navigate through \emph{risk zones} while avoiding obstacles.
Travelling these regions incurs a penalty which is \emph{super-linear} in the traversal time (see Fig.~\ref{fig:example}.
We call the class of problems 
\emph{Risk Aware Motion Planning (RAMP)}
and use a natural cost function which simultaneously optimizes for paths that are both short and reduce consecutive exposure time in the risk zone.

We are motivated by real-world problems involving \emph{risk}, where 
 continuous exposure is much worse than intermittent exposure.
Examples include pursuit-evasion
where sneaking in and out of cover is the preferred strategy,
and visibility planning where the agent must ensure that
an observer or operator is minimally occluded. 

In its general form, our problem can be seen as an instance of the \emph{motion-planning problem}~\cite{L06,CBHKKLT05}
which is known to be PSPACE-Hard~\cite{R79}. 
Thus, in high-dimensional spaces, a natural approach is to follow the \emph{sampling-based paradigm} by computing a discrete graph which is then traversed by a \emph{path-finding} algorithm.
Standard path-finding algorithms such as Dijkstra~\cite{D59} and A*~\cite{HNR68} cannot be used as optimal plans do not posses optimal substructure.
Having said that, we recently suggested efficient path-planning algorithms~\cite{SHS17}.

When restricting the planning domain to the two-dimensional plane it is not clear whether the problem is computationally hard or not.
It is well known that planning for shortest paths in the plane amid polygonal obstacles can be computed in $O(n \log n)$ time, where~$n$ is the  complexity of the obstacles (see~\cite{M04} for a survey).
When computing shortest paths amid 
polyhedral obstacles in $\R^3$,
or in $\R^2$ when there are constraints on the curvature of the path,
the problem becomes NP-Hard~\cite{CR87,KKP11}.
Furthermore, 
the Weighted Region Shortest Path Problem, 
which is closely related to our problem~\cite{MP91},
is unsolvable in the Algebraic Computation Model over the Rational Numbers~\cite{DGMOS14}.
If our problem is computationally hard, as we conjecture,  then a reduction, possibly along the lines of~\cite{CR87} should be provided together with an approximation algorithm.
Here, a possible approach would be to sample the boundary of \Crisk, similar to~\cite{AFS16}.
For a survey of planning algorithms low dimensions, see, e.g.,~\cite{HSS16}

\section{Problem formulation}
\label{sec:problem_formulation}

%
\begin{figure}[tb]
  \centering
  	\includegraphics[height = 4.5cm ]{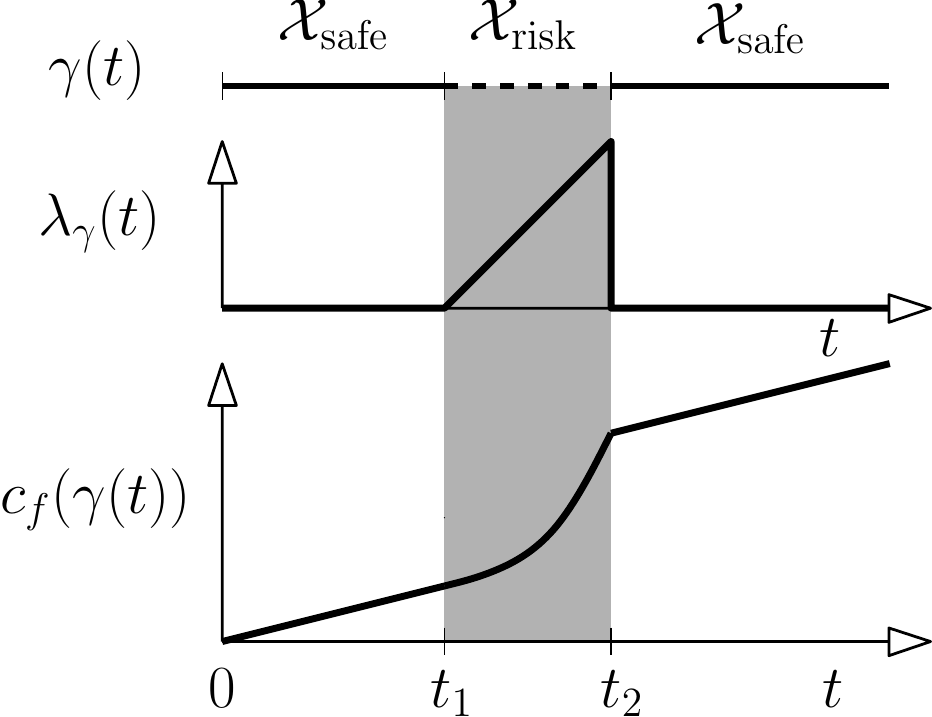}
  \caption{
    \captionstyle
  	Relation between a trajectory~$\gamma(t)$ (top), recent exposure time~$\lambda_\gamma(t)$ (middle) and cost~$c_f(\gamma(t))$ (bottom) as a function of time.
  	In $t \in [0,t_1]$, $\gamma$ stays in~\Csafe, hence~$\lambda_\gamma(t)=0$ and the cost grows linearly with time.
  	At~$t=t_1$, $\gamma$ enters~\Crisk,~$\lambda_\gamma(t)$ grows linearly and the cost grows super-linearly. 
  	At $t=t_2$, $\gamma$ leaves~\Crisk,~$\lambda_\gamma(t)=0$ and the cost returns to growing linearly.   	
  	}
   	\label{fig:cost}
\end{figure}

Let 
$\calP =  \{ P_1, \ldots P_{m} \}$
be a set of simple pairwise interior-disjoint polygons having n vertices in total.
We subdivide~$\calP$ into the disjoint sets $\calP_{\text{obs}}$ and $\calP_{\text{risk}}$ which will be used to define the obstacle region~\Cforb and the risk region~$\Crisk$, respectively.
Roughly speaking, these regions are considered to be open sets. However we do not wish to consider points on the boundary of \Cforb and \Crisk as points out of the risk region which are collision free.
This is captured by the following definition:
The \emph{forbidden} region 
$\Cforb  = \text{int}(\calP_{\text{obs}})$ 
is the set of all points in the interior of $\calP_{\text{obs}}$.
The \emph{risk} region
$\Crisk  = \text{int}(\calP_{\text{risk}}) \cup \left( 
\calP_{\text{obs}} \cap \calP_{\text{risk}} \right)$ 
is the set of all points in the interior of
$\calP_{\text{risk}}$ and all points that lie on the border of
$\calP_{\text{risk}}$ and $\calP_{\text{obs}}$.
Finally, the \emph{risk-free} region is defined as 
$\Csafe = \R^2 \setminus \left( \Crisk \cup \Cforb\right)$.

A trajectory $\gamma: [0,T_\gamma] \rightarrow \R^2$ is a continuous mapping between time and points.
We say that $\gamma$ is \emph{collision free} if $\forall t~\gamma(t) \in \Csafe \cup \Crisk$. 
The image of a trajectory is called a path.
Given a trajectory $\gamma$, 
and some time $t \in [0, T_\gamma]$,
let $t' \leq t$ be the latest time such that $\gamma(t') \in \Csafe$.
Notice that if $\gamma(t) \in \Csafe$ then $t'=t$.
We define the \emph{current exposure time} of $\gamma$ at $t$ as $\lambda_\gamma(t) = t - t'$.
Namely, if $\gamma(t) \in \Crisk$ then $\lambda_\gamma(t)$ is the time passed since $\gamma$ last entered \Crisk.
If $\gamma(t) \in \Csafe$ then $\lambda_\gamma(t)=0$.

We are now ready to define our cost function.
Let 
$\gamma$ be a trajectory and 
$f(x)$ any function such that 
$f(x) = \omega(x)$ and~$f(0) = 1$.
The cost of $\gamma$, denoted by~$c_f (\gamma)$ is defined as 
\begin{equation}
\label{eq:cost}
 c_f (\gamma)
 =
 \int_{t \in [0,T_\gamma]}
 	f(\lambda_{\gamma}(t)) |\dot{\gamma(t)}| dt.
\end{equation}
Eq.~\ref{eq:cost} penalizes continuous exposure to risk in a super-linear fashion (hence the requirement that $f(x) = \omega(x)$). 
As~$f(0) = 1$, the cost of traversing the risk-free region is simply path length.
See Fig.~\ref{fig:cost} for a conceptual visualization of the current exposure time and our cost function.

Equipped with 
our cost function we can formally state the 
risk-aware motion-planning problem:

\vspace{2mm}

\noindent
\emph{Planar Risk-aware motion-planning problem (pRAMP)}
Given the tuple 
$(\Csafe, \Crisk, \Cforb, s, g, f)$,
where $s,  g \in \Csafe$ are 
start and goal points,
compute
$\underset{\gamma \in \Gamma}{\arg\min} \ c_f (\gamma)$
with~$\Gamma$ the set of all collision-free trajectories connecting $s$ and
$g$

We defined our problem to be as general as possible.
However, to simplify the discussion,
we assume that the robot is moving in constant speed and we use $f(x) = e^x$.
Thus, we can re-write Eq.~\ref{eq:cost} as
\begin{equation}
\label{eq:e-cost}
 c (\gamma)
 =
 \int_{t \in [0,T_\gamma]}
 	e^{\lambda_{\gamma}(t)} dt.
\end{equation}
Using the assumption that the robot is moving in constant speed, we can use the terms 
duration of a trajectory and path length interchangeably
(here we measure path length as the Euclidean distance).
Further exploiting this assumption and by a slight abuse of notation we can also use Eq.~\ref{eq:e-cost} to define the cost of a path (and not of a trajectory).
For different properties of this cost function, the reader is referred to~\cite{SHS17}.

\section{Discussion and open questions}
\label{sec:disc}
\subsection{Hardness}
When considering the complexity of a planning problem, one needs to consider both the algebraic complexity and the combinatorial complexity. 
If we use the Algebraic Computation Model over the Rational Numbers (ACM$\Q$), then we conjecture that the problem is unsolvable.
A proof may follow the lines taken in~\cite{DGMOS14} for the 
Weighted Region Shortest Path Problem.

Assessing the combinatorial complexity of our problem, defined analogously to the number of ``edge sequences'' is not as straightforward.
Several hardness results for planning problems use reductions from 4CNF-satisfiability~\cite{AKY03,KKP11}.
The proofs use the idea of ``path encoding'' which 
involves constructing an environment that admits an exponential number of distinct shortest paths between $s$ and $t$. 
Each path is associated with a truth assignment of a given formula $\Phi$.
Then, the environment is augmented with additional obstacles that block  every  path whose associated truth assignment does not satisfy the formula $\Phi$.
The underlying problem with using this approach is that in the plane it depends heavily on
the fact that a minimal-cost paths can self-intersect, which is not the case in our setting.

\subsection{Approximation algorithm}
Assuming that the problem is computationally hard, we seek an approximation algorithm such that given some~$\varepsilon$ returns a path whose cost is at most $1+\varepsilon$ the cost of the optimal path in time polynomial in $n$ and $\varepsilon$.
A natural approach would be to sample densely along the boundary of~$\Crisk$ and compute the visibility graph defined over the sampled points and the vertices in $\calP$. 
A minimal-cost path may then be computed in polynomial time~\cite{SHS16}.
However, the running time of this algorithm also depends on the \emph{length} of the edges of polygons in $\calP$ (see similar approach and analysis in~\cite{WBH08}).

We believe that a possible approach would be to sample the boundary of \Crisk more carefully, similar to~\cite{AFS16}.


\end{document}

%% file: macros.tex
\newcommand{\calX}{\ensuremath{\mathcal{X}}\xspace}

\newcommand{\calP}{\ensuremath{\mathcal{P}}\xspace}

\newcommand{\R}{\mathbb{R}}

\newcommand{\Q}{\mathbb{Q}}

\newcommand{\Cforb}{\ensuremath{\calX_{\rm obs}}\xspace}
\newcommand{\Crisk}{\ensuremath{\calX_{\rm risk}}\xspace}
\newcommand{\Csafe}{\ensuremath{\calX_{\rm safe}}\xspace}

\newcommand{\ignore}[1]{}

\newboolean{ShowAuthors}
\setboolean{ShowAuthors}{false}
\def\blind#1{
\ifthenelse{\boolean {ShowAuthors}}{#1}{}
}

\newcommand{\captionstyle}{\sf \footnotesize }